\documentclass[sigconf]{acmart}

\usepackage{booktabs} 
\usepackage{multirow}
\usepackage{subfigure}
\usepackage{balance}

\copyrightyear{2019} 
\acmYear{2019} 
\setcopyright{acmcopyright}
\acmConference[GLSVLSI '19]{Great Lakes Symposium on VLSI 2019}{May 9--11, 2019}{Tysons Corner, VA, USA}
\acmBooktitle{Great Lakes Symposium on VLSI 2019 (GLSVLSI '19), May 9--11, 2019, Tysons Corner, VA, USA}
\acmPrice{15.00}
\acmDOI{10.1145/3299874.3318037}
\acmISBN{978-1-4503-6252-8/19/05}
\renewcommand\footnotetextcopyrightpermission[1]{} 

\begin{document}
\title{AQuRate: MRAM-based Stochastic Oscillator for Adaptive\\Quantization Rate Sampling of Sparse Signals}

 \author{Soheil Salehi, Ramtin Zand, Alireza Zaeemzadeh, Nazanin Rahnavard, Ronald F. DeMara}
 \affiliation{%
   \institution{Department of Electrical and Computer Engineering, University of Central Florida, Orlando, FL, 32816 USA}
}





\renewcommand{\shortauthors}{S. Salehi et al.}

\begin{abstract}
Recently, the promising aspects of compressive sensing have inspired new circuit-level approaches for their efficient realization within the literature. However, most of these recent advances involving novel sampling techniques have been proposed without considering hardware and signal constraints. Additionally, traditional hardware designs for generating non-uniform sampling clock incur large area overhead and power dissipation. Herein, we propose a novel non-uniform clock generator called Adaptive Quantization Rate (AQR) generator using Magnetic Random Access Memory (MRAM)-based stochastic oscillator devices. Our proposed AQR generator provides $\sim25$-fold reduction in area, on average, while offering $\sim6$-fold reduced power dissipation, on average, compared to the state-of-the-art non-uniform clock generators.
\end{abstract}

%
%



\maketitle
\section{Introduction}
\label{sec:introduction}
Recently, non-uniform sampling approaches such as Compressive Sensing (CS) have been proposed to reduce the energy consumption of sampling operation by reducing number of samples in each frame, reduce required storage to save the sampled data, and reduce the data transmission due to lower number of samples taken \cite{Sarvotham2006MeasurementsTheory,Zaeemzadeh2017AdaptiveSignals,Zaeemzadeh2018ARecovery}. Additionally, event-driven sampling, such as level-crossing sampling, has been widely adopted as a promising CS technique to maximize the performance of sampling operation while reducing energy consumption \cite{Wu2017AFilter}. Furthermore, CS techniques are utilized to sample spectrally sparse wide-band signals close to their information rate rather than their Nyquist rate, which can be a challenge using conventional uniform sampling techniques due to the high cost of the hardware that is capable of performing the sampling operation at a high Nyquist rate. 

Despite all the benefits that CS techniques offer, they are typically realized oblivious to the hardware limitations such as energy, bandwidth, and battery capacity. Additionally, signal-dependent constraints such as sparsity and noise level are ignored while studying the quantization rate and resolution trade-off. The aforementioned hardware-dependent and signal-dependent constraints alter during the sampling operation. Thus, an adaptive quantization rate and resolution optimization circuitry is required to maximize sampling performance while minimizing the number of samples to reduce energy consumption, data transmission, and storage. Adaptive quantization rate and resolution sampling might be readily achieved from the algorithm perspective, however it requires a hardware platform that is capable of real-time adaptation according to certain signal behavior such as sparsity rate. Recently, an adaptive optimization of the quantization rate and resolution during signal acquisition has been investigated in \cite{Salehi2018Energy-AwareDevices}. 


Previous works on adaptive quantization rate and resolution ADCs have been implemented using Complementary Metal Oxide Semiconductor (CMOS) technology and considering a low-pass signal model \cite{Bellasi2014ACMOS,Wu2017AFilter}. Herein, we propose an spin-based Adaptive quantization rate (AQR) generator circuit that considers the signal dependent constraint as well as hardware limitations. The proposed AQR generator circuit utilized Magnetic Random Access Memory (MRAM)-based stochastic oscillator devices, which offer miniaturization and significant energy savings \cite{camsari2017implementing}. 


\section{Background and Related Work}
\label{sec:related-work}

Recently researchers have achieved significant performance improvements using sparse signal recovery techniques. Spectrally sparse signals are utilized in many applications such as frequency hopping communications, musical audio signals, cognitive radio networks, and radar/sonar imaging systems 
\cite{Salehi2018Energy-AwareDevices}. Additionally, a major challenge in spectrum sensing is that in most cases, the sparse components of the signal are spread over a wide-band spectrum and need to be acquired without prior knowledge of their frequencies. Moreover, spectrum-aware communication networks require Radio Frequency and mixed-signal hardware architectures that can achieve very wide-band but energy-efficient spectrum sensing \cite{Salehi2018Energy-AwareDevices}.


The cornerstone to achieving CS approaches and non-uniform sampling techniques is the utilization of an asynchronous pseudo-random clock generator, usually referred to as non-uniform clock generator, which is consisted of a Linear Feedback Shift Register (LFSR) that selects a clock signal at random from a series of ring oscillators with different frequency and phases \cite{Lee2017ASampling,Osama2016DesignApplications,Bhatti2007StandardSignals,Bellasi2014ACMOS,Kose2008Pseudo-randomIntegrity}. In most cases, these circuits require a large number of CMOS transistors and incur significant area overhead and power dissipation. Recently, a novel approach for generating the non-uniform clock using VCMA-MTJ devices is proposed in \cite{Lee2017ASampling} and the authors have shown that their proposed design can achieve significant area and power dissipation reduction compared to the previous CMOS-based pseudo-random clock generators. However, the authors in \cite{Lee2017ASampling} considered the frequency of the signal in order to generate the sampling clock, which limits the bandwidth and in case of spectrally sparse signals, where no prior knowledge of frequency is available, their proposed approach will face challenges. Herein, we consider the sparsity rate of the signal to generate the sampling clock. This will minimize the number of samples and results in more energy savings. Furthermore, our proposed design has reduced complexity compared to other designs proposed in the literature due to significant reduction in the CMOS circuit elements. 

\section{Adaptive quantization rate Generator}
\label{sec:aqr}

\subsection{MRAM-based Stochastic Device as a Building Block for AQR generator}
\label{subsec:p-bit}

\begin{figure}[!t]
\centering
\includegraphics[width=2.0in]{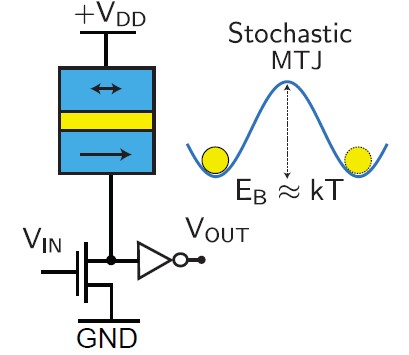}
\caption{The building block of the proposed spin-based AQR generator \cite{camsari2017implementing}.}
\label{fig:pbit}
\end{figure}

In this section, we show that a recently proposed building block with embedded MRAM technology can enable the hardware realization of an AQR generator. The structure of the MRAM-based stochastic device is shown in Fig. \ref{fig:pbit}, which includes a magnetic tunnel junction (MTJ) that is a 2-terminal device with two different resistive levels depending on the orientation of its ferromagnetic (FM) layers, called \textit{fixed layer} and \textit{free layer}. The fixed layer is designed to have a fixed magnetic orientation, while the magnetization orientation of the free layer can be switched. In MRAM-based memory devices, a thermally-stable nanomagnet with a large energy barrier with respect to the thermal energy (kT) is utilized for free layer so that the fixed layer can function as a non-volatile memory. In recent years the use of superparamagnetic MTJs that are not thermally stable have been experimentally and theoretically investigated in search of functional spintronic devices \cite{locatelli2014noise,choi2014magnetic,fukushima2014spin,sutton2017intrinsic,debashis2016experimental,camsari2017stochastic,zandRDBN,zand2018composable}. 

In this paper, we use an MRAM device with a low energy-barrier nanomagnet ($E_B \ll 40kT$), which is thermally unstable \cite{camsari2017implementing}. The resistance of an MTJ with such a low energy barrier nanomagnet randomly fluctuates between high ($R_{AP}$) and low resistance states ($R_{P}$). This creates a fluctuating output voltage at the drain of the NMOS transistor, which can be amplified by an inverter circuit to produce a stochastic output that can be modulated by the input voltage. In particular, the output voltage at the drain of the NMOS transistor can be shorted to the ground by reducing its drain-source resistance ($r_{ds}$) through increasing the input voltage ($V_{IN}$), or it can be near $V_{DD}$ by increasing the $r_{ds}$ through decreasing $V_{IN}$. The device operation can be comprehended by considering the MTJ conductance \cite{camsari2017implementing}:
\begin{equation}
G_{MTJ}  = G_0 \left[ 1 + m_z \frac{TMR}{(2+TMR)}\right]
\label{eq:mtj}
\end{equation}
where $m_z$ is the free layer magnetization that is stochastically fluctuating due to the thermal noise, $G_0$ is the average MTJ conductance, $(G_P + G_{AP})/2$, and $TMR$ is the tunneling magnetoresistance ratio. The drain voltage can be expressed as: 
\begin{equation}
V_{DRAIN}/V_{DD} = \frac{( 2+TMR)+ TMR \ m_z}{(2+TMR)(1+\alpha)+TMR \ m_z}
\label{eq:drain}
\end{equation}
where $\alpha$ is the ratio of the transistor conductance ($G_T$) to the average MTJ conductance ($G_0$). The maximum fluctuations at the drain occurs when $\alpha \approx 1$, thus the MTJ resistance is approximately equal to the NMOS resistance when $V_{IN}=0.5 V_{DD}$. Since the drain voltage fluctuations are in the order of hundreds of mV for typical TMR values, an additional inverter is used to amplify the noise to produce output voltages ranging from 0 to $V_{DD}$.

\subsection{AQR Generator Circuit}
\label{subsec:aqr_circuit}

\begin{figure}[!t]
    \centering
    \includegraphics[width=3.3in]{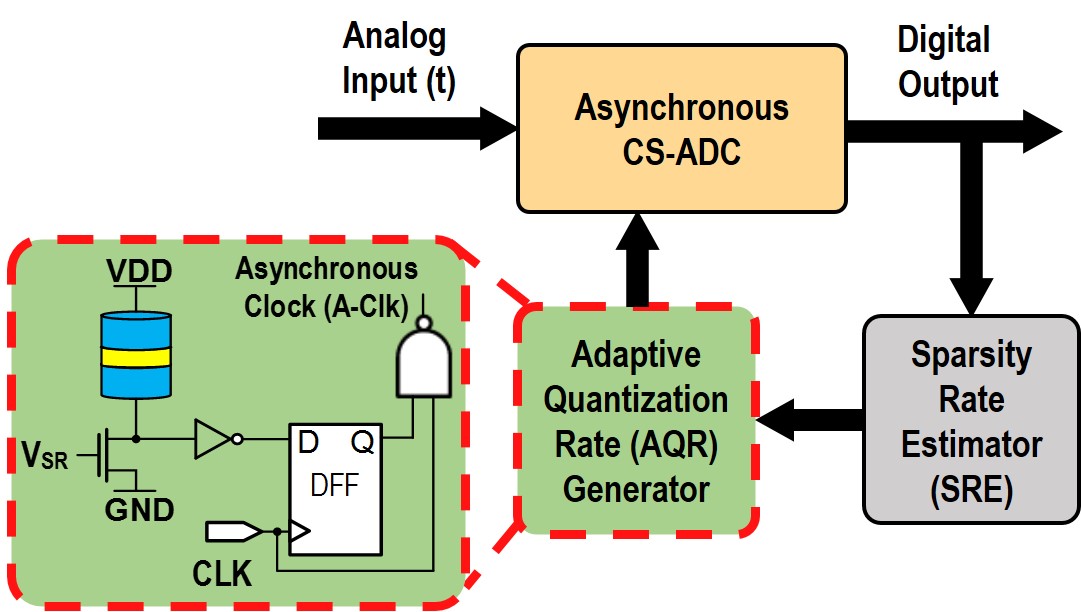}
    \caption{Integration of AQR generator circuit within the Compressive Sensing ADC (CS-ADC) system design.}
    \label{fig:AQR-system}
\end{figure}

To realize an effective hybrid emerging device and CMOS circuit, one useful approach can be to consider stochastic and deterministic attributes separately. For instance, Fig. \ref{fig:AQR-system} depicts the proposed AQR generator circuit wherein a $2$-terminal MTJ realizes stochastic behavior to provide the non-uniform clock generation capability. 

The quantized Sparsity Rate Estimator (SRE) module shown in Fig. \ref{fig:AQR-system} estimates the sparsity rate of the digital output bit-stream by estimating the sparse spectral components of the digital output using an iterative algorithm. Recently, rapid and optimized sparse component estimation method is proposed in \cite{Salehi2018Energy-AwareDevices}. In the approach proposed in \cite{Salehi2018Energy-AwareDevices} in order to minimize the computational complexity of the sparse component estimation, an sliding window approach is utilized and the algorithm operates only one iteration on each frame of the input by utilizing the previous estimate as an initial value. This will result in gradual convergence of the sparse components to the actual values across iterations. These algorithms can be employed to find the sparsity rate of the signal. In most cases, sparsity rate of analog signals, which can be described as the number of non-zero elements in divided by the total number of elements the sparse representation of the signal, is between $5\%$ to $15\%$ in many applications including those targeted herein.

When the SRE module estimates the sparsity rate of the signal based on the digital output of the previous frame, it will then generate a voltage level according to that sparsity rate of the input analog signal. This voltage, referred to as $V_{SR}$, will be applied to the gate of the NMOS transistor shown in Fig. \ref{fig:AQR-system} and results in an stochastic bit-stream generated by the MRAM-based stochastic oscillator device. The stochastic bit-stream output generated by the MRAM-based stochastic oscillator device will be forwarded to the D-Flip-Flop (D-FF) as shown in Fig. \ref{fig:AQR-system} and the result of the $2$-input NAND gate between the output of the D-FF and the actual clock of the circuit will generate the required quantization rate to be used for the following frame of the signal acquisition, referred to as Asynchronous Clock (A-Clk) in Fig. \ref{fig:AQR-system}. Additionally, the SRE module can also used by the recovery algorithms to efficiently recover the sampled signal \cite{Salehi2018Energy-AwareDevices}. Additionally, the A-Clk will be forwarded to the sparse recovery algorithm to provide necessary information about the samples taken from the signal to assist with the signal reconstruction. 

\begin{figure}[!t]
\centering
\includegraphics[width=3.2in]{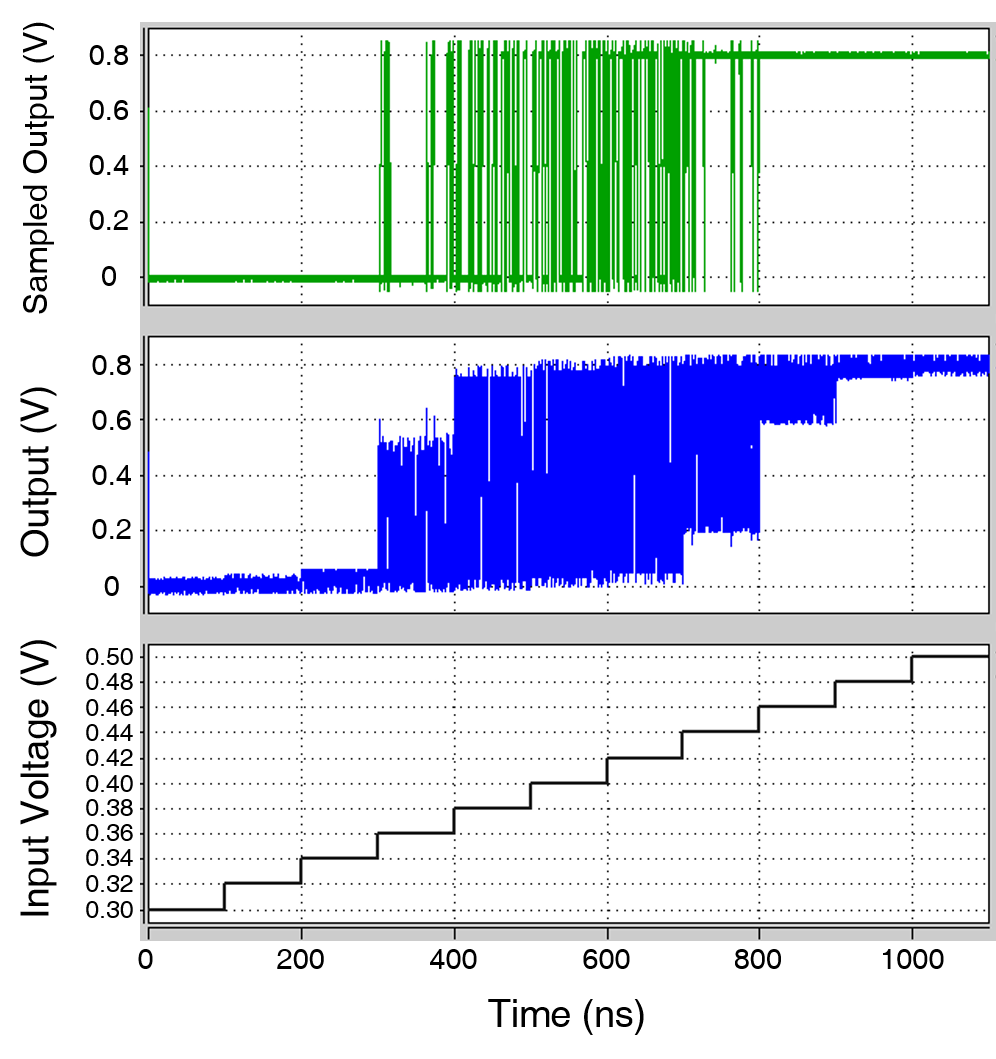}
\caption{The sampled output of the stochastic MRAM-based building block for AQR generator for various input voltages.}
\label{fig:sampleout}
\end{figure}

To obtain the relation between the output probability of the stochastic MRAM-based AQR generator and its input voltage, we have applied an input pulse that its amplitude starts from $GND$ and is increased by $200$mV every $100$ns until it reaches $V_{DD}$. The output of the building block is sampled with a $1$GHz clock frequency using a D-FF circuit, as shown in Fig. \ref{fig:sampleout}. 


\section{Simulation Results}
\label{sec:results}

In order to evaluate and validate the behavior and functionality of the proposed AQR generator circuit, SPICE and MATLAB simulations were performed. We have utilized the $14$nm High Performance FinFET Predictive Technology Model (PTM) \cite{ASU14nmHttp://ptm.asu.edu/} as well as the MRAM-based stochastic oscillator device model and parameters represented in \cite{camsari2017implementing} to implement and evaluate the proposed AQR generator circuit.


According to our results, AQR provides significant power dissipation and area reductions compared to the state-of-the-art nonuniform clock generators listed in Table \ref{tab:compare} \cite{Lee2017ASampling,Osama2016DesignApplications,Bhatti2007StandardSignals,Bellasi2014ACMOS}. According to our simulation results, power dissipation of the proposed AQR generator circuit is $22.64\mu$W on average. With respect to area utilization, our proposed AQR design requires only $23$ FinFET transistors, which attains a significant reduction in the transistor count and complexity of the non-uniform clock generator circuit present in state-of-the-art designs \cite{Lee2017ASampling,Osama2016DesignApplications,Bhatti2007StandardSignals,Bellasi2014ACMOS}. Thus, AQR avoids high transistor counts while making it unnecessary to use of large LFSR circuits that contain numerous D-FFs as well as several logic gates and multiplexers. For a more equitable comparison in terms of area and power dissipation, we have derived (\ref{eq:power_scaled}) and (\ref{eq:area_scaled}) considering General Scaling method \cite{Stillmaker2017Scaling7nm} to normalize the power dissipation and area of the designs listed in Table \ref{tab:compare}. Based on the General Scaling method, voltage and area scale at different rate of $U$ and $S$, respectively. Thus, the power dissipation is scaled with respect to $1/U^2$ and area per device is scaled according to $1/S^2$ \cite{Stillmaker2017Scaling7nm}.

\begin{equation}
\label{eq:power_scaled}
    \centering
    \begin{aligned}
    Power_{norm} = & \frac{Power_x}{Power_{AQR}}\times(\frac{1}{U})^2\\= &
    \frac{Power_x}{Power_{AQR}}\times(\frac{0.8V}{V_{nominal}})^2
    \end{aligned}
\end{equation}
\begin{equation}
\label{eq:area_scaled}
    \centering
    \begin{aligned}
    Area_{norm} = & \frac{Area_x}{Area_{AQR}}\times(\frac{1}{S})^2\\ = &
    \frac{Area_x}{Area_{AQR}}\times(\frac{14nm}{Technology})^2
    \end{aligned}
\end{equation}
\noindent
where, $V_{nominal}$ is the nominal voltage of the technology model, $Technology$ refers to the technology node in nanometers, and subscript $x$ refers to the design that we want to scale its power dissipation and area according to the technology models. According to (\ref{eq:power_scaled}) and (\ref{eq:area_scaled}), AQR provides power dissipation reduction up to one-order-of-magnitude compared to the state-of-the-art nonuniform clock generators as listed in Table \ref{tab:compare}. Additionally, AQR offers up to one-orders-of-magnitude area reduction compared to the designs provided in Table \ref{tab:compare} using the scaling comparison trends accepted in the literature.



\begin{figure}[!t]
    \centering
    \includegraphics[width=3.1in]{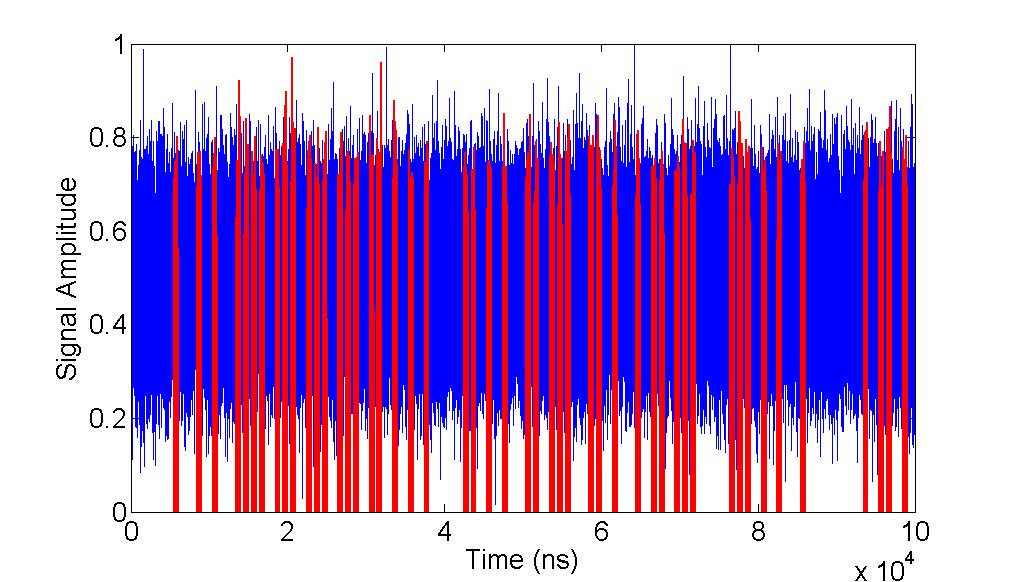}
    \caption{Sampling an analog signal with sparsity rate of $5\%$ using AQR generator. Blue represents the signal and Red represents the samples taken using the AQR generator.}
    \label{fig:outprob}
\end{figure}

\begin{table}[!t]
\centering
\caption{Comparison with recently proposed non-uniform clock generator designs}
\label{tab:compare}
\setlength{\tabcolsep}{0.25em}
\begin{tabular}{c c c c}
\hline
\textbf{Design} &\textbf{Technology (V$_{nominal}$)} &\textbf{Power$_{norm}$} &\textbf{Area$_{norm}$}\\\hline\hline
\cite{Lee2017ASampling} &$65$nm ($1.1$V) &$\sim1\times$ &$\sim1\times$\\

\cite{Osama2016DesignApplications} &$65$nm ($1.1$V) &$\sim2\times$ &$\sim21\times$ \\

\cite{Bhatti2007StandardSignals} &$90$nm ($1.2$V) &$\sim2\times$ &$\sim51\times$ \\

\cite{Bellasi2014ACMOS} &$28$nm ($1.0$V) &$\sim18\times$ &N/A \\

This Work &$14$nm ($0.8$V) &$1\times$ &$1\times$ \\\hline

\end{tabular}
\end{table}
As described in Section \ref{subsec:aqr_circuit}, sparsity rate of analog signals is usually within the range of $5\%-15\%$. Fig. \ref{fig:outprob} depicts an example output of the AQR generator for sampling of a sparse signal with $5\%$ sparsity rate. Moreover, we have embedded our proposed AQR generator within CS recovery algorithms called Orthogonal Matching Pursuit (OMP) and Compressive Sampling Matching Pursuit (CoSaMP) \cite{2009CoSaMP:Samples} in order to evaluate the architectural simulation results and in order to recover the signal from the samples taken using the AQR generator. According to the results, the mean normalized errors of the reconstruction of the signals with $5\%$, $10\%$, and $15\%$ sparsity rates using OMP are $0.0504$, $0.0446$, and $0.0252$, respectively. Moreoever, the mean normalized errors of the reconstruction of the signals with $5\%$, $10\%$, and $15\%$ sparsity rates using CoSaMP are $0.0487$, $0.0304$, and $0.0245$, respectively. 

\section{Conclusions}
\label{sec:conclusion}
We have devised a novel non-uniform clock generator called Adaptive quantization rate (AQR) generator using MRAM-based stochastic oscillator devices. Our proposed AQR generator considers signal constraints, such as sparsity rate, as well as hardware constraints, such as area and power dissipation, in order to generate the non-uniform clock for the asynchronous CS-ADC.
Compared to similar non-uniform clock generators presented in the literature, AQR generator provides significant area reduction of $\sim25$-fold on average, while achieving power dissipation reduction of $\sim6$-fold, on average. 

\section*{Acknowledgement}
This work was supported in part by the Center for Probabilistic Spin Logic for Low-Energy Boolean and Non-Boolean Computing (CAPSL), one of the Nanoelectronic Computing Research (nCORE) Centers as task 2759.006, a Semiconductor Research Corporation (SRC) program sponsored by the NSF through CCF 1739635, and by NSF through ECCS 1810256.
\balance

\bibliographystyle{ACM-Reference-Format}
\bibliography{soheil.bib,ramtin.bib}
\end{document}